# Magnetoresistive behaviour of ternary Cu-based materials processed by high-pressure torsion

M Kasalo[1], S Wurster[1], M Stückler[1], M Zawodzki[1], L Weissitsch[1], R Pippan[1] and A Bachmaier[1]

[1] Erich Schmid Institute of Materials Science of the Austrian Academy of Sciences, Jahnstrasse 12, 8700 Leoben, Austria

**Abstract.** Severe plastic deformation using high-pressure torsion of ternary Cu-based materials (CuFeCo and CuFeNi) was used to fabricate bulk samples with a nanocrystalline microstructure. The goal was to produce materials featuring the granular giant magnetoresistance effect, requiring interfaces between ferro- and nonmagnetic materials. This magnetic effect was found for both ternary systems; adequate subsequent annealing had a positive influence. The as-deformed states, as well as microstructural changes upon thermal treatments, were studied using scanning electron microscopy and X-ray diffraction measurements. Deducing from electron microscopy, a single-phase structure was observed for all as-deformed samples, indicating the formation of a supersaturated solid solution. However, judging from the presence of the granular giant-magnetoresistive effect, small ferromagnetic particles have to be present. The highest drop in room temperature resistivity (2.45% at 1790 kA/m) was found in $Cu_{62}Fe_{19}Ni_{19}$ after annealing for 1 h at 400 °C. Combining the results of classical microstructural studies and magnetic measurements, insights into the evolution of ferromagnetic particles are accessible.

## 1. Introduction
The change of the electrical resistance in an external magnetic field is termed magnetoresistance (MR) and depending on the material, different MR-effects can occur. Fert et al. and Grünberg et al. independently showed that the change in MR can be huge, for instance ~50% were found in alternating thin layers of Cr and Fe at low temperatures. Due to the large changes in resistance, it was labelled giant magnetoresistance (GMR) [1,2]. The GMR effect is distinguished from the ordinary, anisotropic MR effect by its isotropic behaviour. Currently, research is conducted on e.g. its application for MR-RAM or the detection of influenza A virus by GMR biosensors [3]. Monoclonal antibodies are exchanged for a combination of viral nucleoproteins and magnetic nanoparticles. In the presence of an influenza virus, the magnetic nanoparticles can be bound to the GMR biosensor and the binding rate is proportional to the concentration of the virus. A change in concentration results in a change in resistance, which can be measured in real-time [4].

Alternating ferromagnetic (FM) and nonmagnetic (NM) multi-layered structures are no necessity for GMR. Bulk materials exhibiting small FM particles embedded in a NM matrix can also show a GMR behaviour, called *granular* GMR [5,6]. In general, the evolution of the GMR with increasing magnetic field is connected to the magnetization of the material. For materials featuring the granular GMR, a disadvantage is that the saturation of the hysteresis and thus the full effect of GMR is reached at high magnetic fields. Regarding application, however, 3D-bulk materials show more freedom in application-relevant shaping compared to 2D-multilayered materials.

Typical methods to fabricate granular alloys are mechanical alloying, ion implantation, melt spinning and co-evaporation. It was already demonstrated that materials processed by high-pressure torsion (HPT), a severe plastic deformation technique, also show a GMR effect [7–9]. Fathoming the prospects of HPT together with a proper concatenation of annealing treatments, this work focuses on the granular GMR effect of a variety of ternary Cu-based materials. For the existence of an interface between FM and NM phases, the chosen material systems have to show a pronounced miscibility gap. Fe-Co, due to its increased saturation magnetization, and Fe-Ni, due to its easy processability, were chosen as FM phases and Cu for the NM phase.

## 2. Experimental
The used powders are Cu (99.9% purity, 170+400 mesh, Alfa Aesar), Co (99.998%, -22 mesh, Alfa Aesar Puratronic), Fe (99.9%, -100+200 mesh, MaTeck), Ni (99.9%, -3+6 mesh, Alfa Aesar). The FM volume fraction was chosen to be 20 vol.%, as it is known from the literature that this ratio leads to the highest granular GMR [10]. In addition, samples containing 40 vol.% of FM material were produced. In the first step, the pure elemental metal powders were mixed to the desired composition. In the second step, the powder blends were pre-compacted in Ar atmosphere to obtain a bulk sample. In the last step, the powder compacts were HPT-deformed at room temperature (RT) at a nominal pressure of 5 GPa, using a rotational speed of 1.28 min$^{-1}$ for 100 rotations. Annealing treatments, followed by water quenching, were performed using a conventional furnace. For long-time annealing treatments (100 h) or treatments at the highest temperature (600 °C) the samples were encapsulated in Ar-filled glass tubes.

For RT resistivity measurements within a magnetic field, the experimental setup described in [8,9] was used. Herein, GMR is represented by the difference in resistivity at zero field and highest field in relation to the resistivity at zero field. Two different orientations of current flow in relation to the magnetic field lines are probed. The exact chemical composition of each material system was determined by electron dispersive X-ray spectroscopy (e-Flash, Bruker) attached to a scanning electron microscope (SEM, Magna, Tescan). All compositions of HPT-samples are given in wt.%. Microstructural investigations in different viewing directions (LEO1525, Zeiss) and X-ray diffraction (XRD) measurements (D2-Phaser, Bruker) of all states were performed.

## 3. Results and Discussion

### 3.1. Scanning electron microscopy
As shown in figure 1a, a homogeneous microstructure in the nanocrystalline (NC) regime was reached for $Cu_{78}Fe_{16}Co_6$ after deformation. The low phase-contrast in backscatter mode indicates a good intermixing. The additional annealing treatments for 1 h at 400 °C and 10 min at 500 °C do not increase the grain size significantly. However, an increased number of small, darker particles is observed in figure 1b compared to 1a and 1c. These can be identified as FM particles (Fe and/or Co) due to Z-contrast.

For $Cu_{62}Fe_{19}Ni_{19}$, the as-deformed state and two chosen annealed samples are presented in figure 1d-f. In comparison to the as-deformed state, figure 1d, annealing for 1 h at 300 °C and 400 °C (figure 1e) caused hardly any grain growth. The achieved microstructure is somewhat coarser for higher annealing durations, 100 h at 400 °C, and higher annealing temperatures, 500 °C (figure 1f) and 600 °C. However, a homogeneous NC microstructure was retained in each case, even at the highest annealing temperature. Deduced from the very small changes in grain size, the microstructure of $Cu_{62}Fe_{19}Ni_{19}$ shows remarkable temperature stability up to 600 °C, which corresponds to a homologous temperature $T_{hom,Cu}$ of 0.64. The coarsest microstructure was obtained for the longest annealing duration of 100 h at 400 °C; however, the grain sizes still being in the NC regime.

$Cu_{84}Fe_7Ni_9$ was exposed to the same heat treatments as $Cu_{78}Fe_{16}Co_6$. The microstructure of the as-deformed material is shown in figure 1g. A homogeneous NC microstructure was reached in each state with the subsequent thermal treatments for 1 h at 400 °C (figure 1h) and 10 min at 500 °C (figure 1i) resulting in insignificant grain growth.

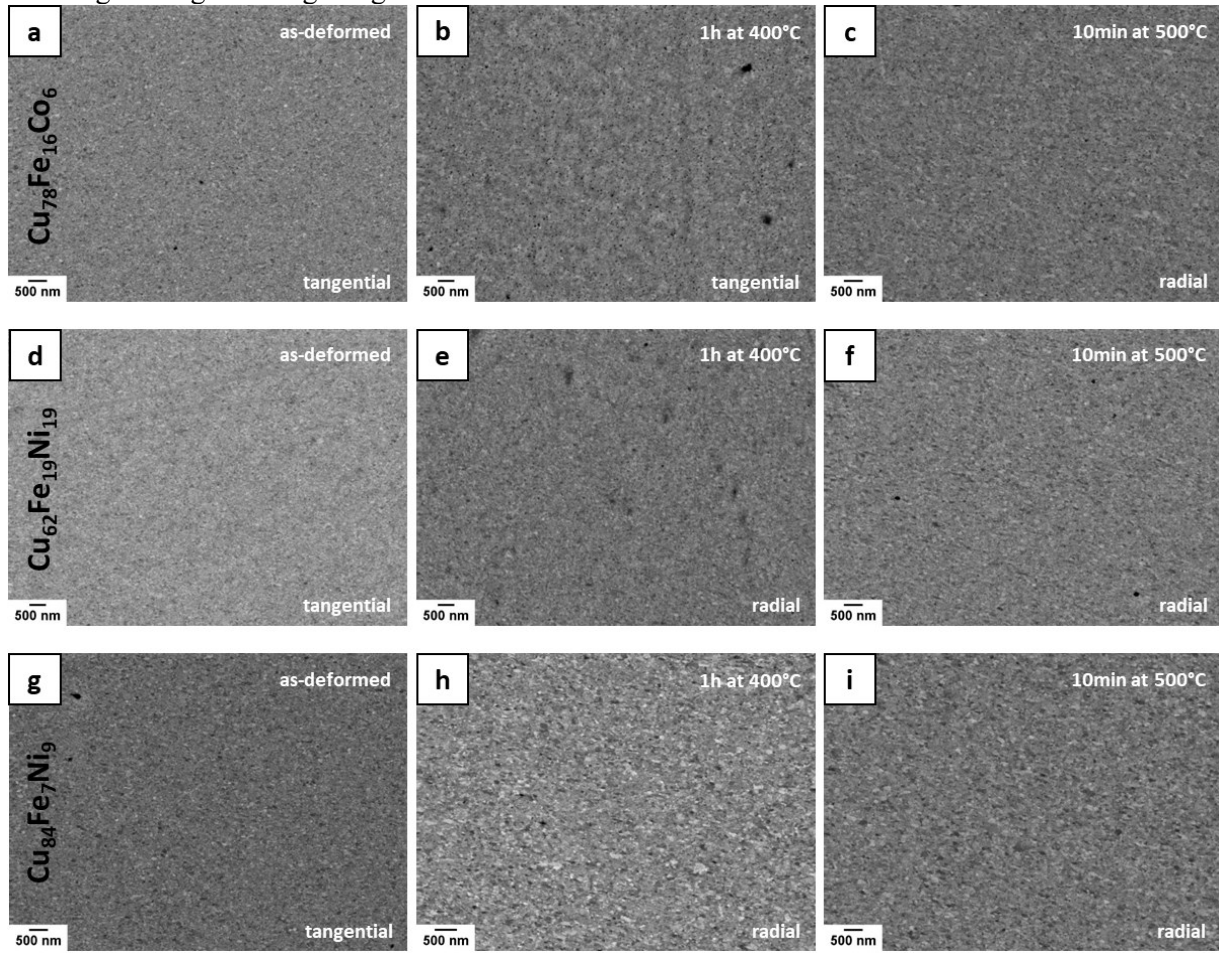

**Figure 1.** SEM micrographs in backscatter mode of all investigated materials in as-deformed and annealed states. All micrographs were taken at a radius of ~3 mm; the viewing direction with respect to the HPT-disc is indicated.

*3.2. XRD*

XRD measurements were performed to study the crystallographic phases of all samples. They were carried out on the samples which were used for electrical measurements; thus, the XRD-investigated spots are located at a radius larger than 1.5 mm from the specimen's center. For $Cu_{78}Fe_{16}Co_6$, no Fe-bcc peaks are present in the as-deformed state (figure 2a, lower part), indicating again the formation of a supersaturated solid solution, although Co-Cu and Fe-Cu are immiscible systems at equilibrium conditions [11]. In case of partial supersaturation, another possible explanation for vanishing Fe-peaks could be their low intensity, as bcc-phase content would be strongly reduced in volume after HPT. After annealing the samples, Fe-bcc peaks occurred for $Cu_{78}Fe_{16}Co_6$, indicating Fe precipitation and grain growth. Furthermore, annealing narrows the full width at half maximum (FWHM), which is in good accordance with the BSE micrographs in figure 1a-c.

Figure 2b, shows XRD results of as-deformed and annealed $Cu_{62}Fe_{19}Ni_{19}$. A single-phase fcc structure close to Cu-fcc was reached for the as-deformed state. No peaks appear at the Fe-bcc and Ni-fcc positions. Therefore, one can assume that Fe and Ni are dissolved in the fcc Cu-matrix. The annealing leads to peak narrowing, especially for 100 h at 400 °C and 1 h at 600 °C. This indicates grain growth

and correlates with the BSE micrographs in figure 1d-f. In addition, the high-temperature stability of the $Cu_{62}Fe_{19}Ni_{19}$ alloy is confirmed, since no phase separation occurs, even after annealing at 1 h at 600 °C.

For $Cu_{84}Fe_7Ni_9$, a single-phase fcc structure in the as-deformed and annealed states is present (figure 2a, upper part). While $Cu_{78}Fe_{16}Co_6$ samples show a phase separation after annealing, CuFeNi-samples do not show this behaviour, though being exposed to the same annealing parameters.

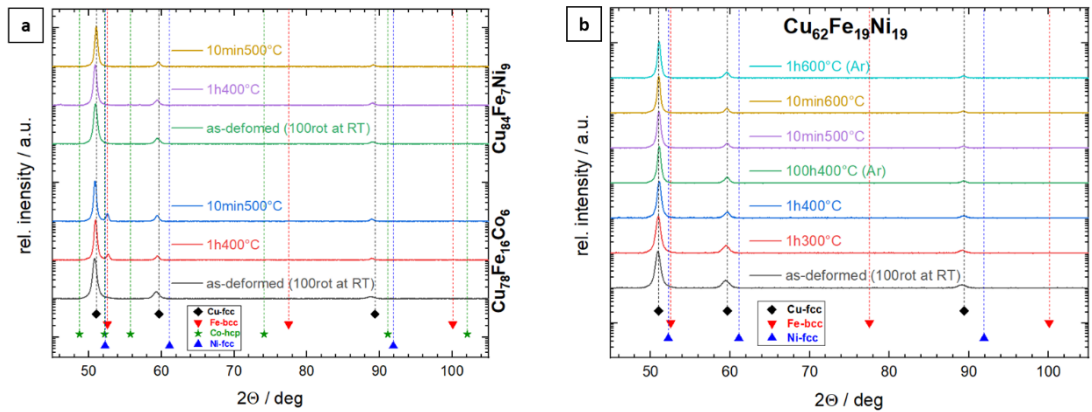

**Figure 2.** XRD patterns of a) $Cu_{78}Fe_{16}Co_6$ and $Cu_{84}Fe_7Ni_9$, and b) $Cu_{62}Fe_{19}Ni_{19}$ for as-deformed and annealed states. Measurements were performed in the axial direction at a radius > 1.5 mm.

*3.3. Magnetoresistive measurements*
In figure 3, the decrease of resistivity with increasing applied magnetic field is shown for all as-deformed and annealed states. $Cu_{78}Fe_{16}Co_6$ exhibits an isotropic GMR behaviour in each state; the linear decrease of perpendicular and parallel measurements is almost identical (figure 3a). At the maximum applied field $H_{max}$ of 1790 kA/m, a GMR of about 1.1% was reached for the as-deformed state. In general, for the occurrence of a granular GMR effect, interfaces between FM and NM phases are required. In SEM- and XRD-data, these interfaces are not clearly evident. However, a granular GMR effect was yet measured. Stückler et. al showed with atom probe tomography measurements that supersaturation of Fe (between 7 wt.% and 25 wt.%) in Cu on the atomic scale can be obtained after HPT deformation. Still, some small Fe-rich particles remain in the Cu-matrix [12]. It is evident from MR results in figure 3 that these interfaces; thus the FM particles, must be present, though the particles are so small that they cannot be detected by low energy XRD and SEM.

For $Cu_{78}Fe_{16}Co_6$, subsequent annealing led to an amplification of the GMR amplitude. The highest GMR of ~1.45% at $H_{max}$ was reached after 1 h at 400 °C. Compared to the literature, a GMR of ~15% was observed for ball-milled $Cu_{80}Fe_{14}Co_6$ powder annealed for 1 h at 415 °C [13]. Furthermore, the annealed material reaches even higher values up to 16% at fields of ~4000 kA/m, where saturation is finally reached [13]. It is important to note that these measurements were carried out at 5 K, where higher GMR values are accessible. As can be seen in figure 3a, no saturation in GMR occurs for $Cu_{78}Fe_{16}Co_6$ at $H_{max}$, even for the annealed samples. This behaviour could be caused by the expected slowly rising hysteresis curve of these materials due to partial supersaturation, i.e. a fraction of FM atoms is diluted in a NM matrix giving rise to a change of the FM long-range interaction [14]. It can be assumed due to literature values [13] that a saturation behaviour of $Cu_{78}Fe_{16}Co_6$ is expected to be reached at much higher applied magnetic fields and/or lower temperatures.

MR measurements of as-deformed and annealed $Cu_{84}Fe_7Ni_9$ are shown in figure 3b. All MR measurements show an isotropic GMR behaviour. For the as-deformed sample, a GMR of 0.6% was reached at $H_{max}$. Both subsequent annealing treatments, 1 h at 400°C and 10 min at 500 °C, lead to an improvement of the GMR amplitude. The obtained GMR of 1.3% at $H_{max}$ is more than two times higher compared to the as-deformed state. Annealing for 10 min at 500 °C results in a maximum GMR of

~2.1%, which is more than three times the value of the as-deformed sample. Furthermore, this state shows a tendency of a saturation behaviour and the GMR values decrease quadratically with increasing field. Thus, the change in resistivity is expected to correlate well with the squared global relative magnetization according to the relationship already described in [5]. In comparison, the same heat treatment (10 min at 500 °C) resulted only in a weak raise of the GMR amplitude in $Cu_{78}Fe_{16}Co_6$ and a reduction of the GMR amplitude in $Cu_{62}Fe_{19}Ni_{19}$ compared to their as-deformed state. Annealing for 1 h at 400 °C had in each case a positive effect regarding the change in resistivity. Referring to the literature [16], GMR values of 8% were found in $Cu_{80}Fe_{10}Ni_{10}$ melt spun ribbons. An even higher GMR amplitude of 19% was achieved by subsequent annealing for 2 h at 400 °C. However, these values were measured at a low temperature of 50 K and a very high field of ~5570 kA/m.

The as-deformed $Cu_{62}Fe_{19}Ni_{19}$ sample (figure 3c) shows a linear drop in resistivity with increasing magnetic field. Demonstrating reproducibility, three samples of the $Cu_{62}Fe_{19}Ni_{19}$ were HPT-deformed and measured. For the measurement in parallel field alignment, they show perfect agreement of GMR values at the highest field of 1.84%, 1.85%, and 1.80%, respectively. Two annealed samples (1 h at 300 °C and 1 h at 400 °C) show a linear decrease of GMR values without any saturation behaviour. An increase of the GMR effect was achieved by annealing for 1 h at 400 °C. At higher annealing temperatures or for longer times (100 h at 400 °C, 10 min at 500 °C, 10 min at 600 °C, and 1 h at 600 °C) the changes in GMR become smaller. In all cases, a reduction of GMR amplitudes was found. A GMR of ~2.4% was observed at similar conditions (RT measurement up to H ~ 1590 kA/m) in magnetron sputtered and spinodal decomposed $Cu_{60}Fe_{20}Ni_{20}$ films [15]. Higher GMR values of 6.5% were reached for higher fields of ~4770 kA/m, whereby the GMR curve does not saturate even for the highest fields. For the presented HPT-deformed material, it can be assumed that increased GMR amplitudes will be achieved in higher applied magnetic fields.

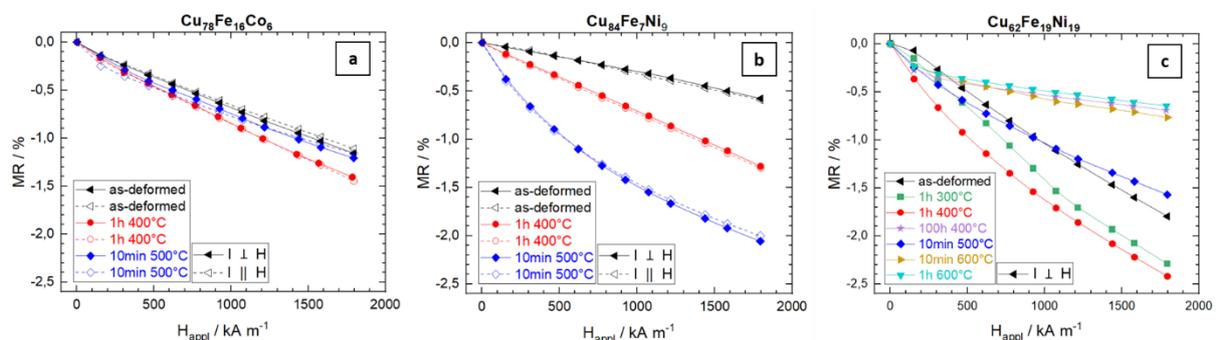

**Figure 3.** Decrease in resistivity with increasing applied magnetic field of a) $Cu_{78}Fe_{16}Co_6$, b) $Cu_{84}Fe_7Ni_9$, and c) of $Cu_{62}Fe_{19}Ni_{19}$ for as-deformed and annealed states. For a) and b) both measurement directions with current flow parallel (open symbols) or perpendicular (filled symbols) to the applied magnetic field are shown. For the sake of clarity, only the perpendicular measurement direction is presented for $Cu_{62}Fe_{19}Ni_{19}$.

## 4. Conclusion

The granular GMR effect was measured in three different ternary Cu-based materials processed by HPT deformation. At RT and the highest achievable magnetic field of 1790 kA/m, $Cu_{62}Fe_{19}Ni_{19}$ shows the highest GMR amplitude of all as-deformed states. Subsequent heat treatments (1 h at 400 °C and 10 min at 500 °C) led to higher GMR amplitudes in $Cu_{78}Fe_{16}Co_6$ and $Cu_{84}Fe_7Ni_9$ samples. Generally, the highest GMR of 2.45% at 1790 kA/m was found in $Cu_{62}Fe_{19}Ni_{19}$ after annealing for 1 h at 400 °C. For this material, a weakening of the effect was observed for high annealing temperatures and long durations. The maximum measured GMR of $Cu_{62}Fe_{19}Ni_{19}$ is somewhat higher than the reported GMR value of spinodal decomposed $Cu_{60}Fe_{20}Ni_{20}$ films at similar measuring conditions [15]. In comparison to the literature, $Cu_{78}Fe_{16}Co_6$ and $Cu_{84}Fe_7Ni_9$ show smaller GMR values. However, no saturation in GMR was observed in any Cu-based composition in as-deformed and annealed states. Higher magnetic fields or lower temperatures are expected to lead to a stronger GMR effect. Furthermore, the microstructure of as-deformed $Cu_{62}Fe_{19}Ni_{19}$ exhibits astonishingly high-temperature stability up to 600 °C, which corresponds to a homologous temperature of 0.64 of Cu. Comparing both material systems, CuFeNi exhibits a higher single-phase stability than CuFeCo.

**Funding:** This project has received funding from the European Research Council (ERC) under the European Union's Horizon 2020 research and innovation programme (Grant No. 757333).